\documentclass[french]{article-hermes}
\usepackage[T1]{fontenc}
\usepackage{graphicx}

\newcommand{\noun}[1]{\textsc{#1}}

\usepackage{babel}
\usepackage{times}

\begin{document}
\title[Contractualisation des ressources]{Contractualisation des
  ressources\\pour le déploiement des composants logiciels}

\date{}

\author{Nicolas Le Sommer} 

\address{Laboratoire Valoria\\ Université de Bretagne Sud\\ Centre de
  recherches Yves Coppens\\ Campus de Tohannic \\ 56000 Vannes, France
  \\ \texttt{Nicolas.Le-Sommer@univ-ubs.fr}}

\resume{Le déploiement de composants logiciels peut s'avérer être une
  tâche complexe et difficile à réaliser dès lors que les composants
  logiciels exhibent des propriétés non-fonctionnelles. En effet, de
  tels composants peuvent ne pas fonctionner de manière convenable si
  certaines de ces propriétés ne peuvent être satisfaites par leur
  plate-forme de déploiement.  Dans cet article, nous nous proposons
  de prendre en compte via une approche contractuelle et réflexive une
  catégorie spécifique des exigences non-fonctionnelles exprimées par
  les composants logiciels, à savoir celles portant sur les ressources
  qui sont nécessaires à leur exécution. }

\motscles{Composants logiciels, ressources, contrats, deploiement} 

\abstract{ Software deployment can turn into a baffling problem when
  the components being deployed exhibit non-functional requirements.
  If the platform on which such components are deployed cannot satisfy
  their non-functional requirements, then they may in turn fail to
  perform satisfactorily. In this paper, we present a contract-based
  approach to take a specific category of non-functional properties
  specified by components into account, that is those that pertain to
  the resources that are necessary for their execution. }


\proceedings{DECOR'04, Déploiement et (Re)Configuration de Logiciels}{211}

\maketitlepage


\section{\label{sec:introduction}Introduction }

De nos jours, les composants logiciels sont devenus des éléments
architecturaux incontournables dans le monde du logiciel. Ils doivent
permettre d'assimiler à terme le développement du logiciel à un simple
processus d'assemblage.  Cet objectif ne pourra être atteint qu'à la
condition de disposer de composants logiciels conçus comme de
véritables unités de déploiement fiables, performantes et parfaitement
décrites tant du point de vue de leurs propriétés fonctionnelles que
de leurs propriétés non-fonction\-nelles.  Les modèles de composants
actuels (e.g. \noun{Corba} Component Model, Entreprise Java Beans,
Fractal, \noun{Com}, OSGi) n'offrent pas aux programmeurs le moyen de
spécifier toutes les propriétés non-fonctionnelles de leurs
composants. Ainsi, les ressources nécessaires à l'exécution des
composants logiciels --qui constituent l'une des principales
catégories de propriétés non-fonctionnelles pouvant être exhibées par
les composants-- sont actuellement ignorées dans ces modèles de
composants. Les composants logiciels n'ont pourtant pas tous les mêmes
besoins vis-à-vis des ressources. Certains composants peuvent
fonctionner avec peu de ressources et avec aucune garantie de
disponibilité de ressources (e.g. des composants réalisant des calculs
élémentaires), alors que d'autres composants vont exiger de disposer
de ressources en quantités importantes et des garanties de
disponibilité vis-à-vis de ces ressources afin de pouvoir fournir à
leurs utilisateurs un certain niveau de qualité de service (e.g. des
composants traitant et affichant des flux vidéos).

Dans cet article, nous nous proposons de prendre en compte les besoins
en ressources des composants logiciels avec une approche contractuelle
et réflexive. Cette approche vise à permettre aux composants logiciels
d'indiquer à leur environnement de déploiement qu'ils n'utiliseront
pas plus de ressources que celles dont ils font explicitement la
demande, et à leur permettre de bénéficier en retour d'un certain
service de la part de cet environnement d'accueil vis-à-vis de la
disponibilité des ressources demandées. Cette approche a également
pour objectif d'offrir aux composants logiciels le moyen de raisonner
sur leurs conditions d'exécution à travers les contrats qu'ils
souscrivent avec leur environnement de déploiement, de pouvoir
modifier ces conditions d'exécution en renégociant dynamiquement les
modalités de leurs contrats, et éventuellement de pouvoir adapter leur
comportement en fonction des contrats négociés. Nous montrons
également dans cet article comment cette approche a été mise en \oe
uvre au sein d'une plate-forme de déploiement expérimentale baptisée
\noun{Jamus}.

La suite de cet article s'articule de la manière suivante. Le
paragraphe~\ref{sec:overview} offre une vue d'ensemble de notre
approche. Le paragraphe~\ref{sec:contrats} propose une modélisation
orientée objet des contrats portant sur les conditions d'accès aux
ressources. Le paragraphe~\ref{sec:jamus} présente les mécanismes qui
permettent de gérer ces contrats de manière dynamique, ainsi que la
mise en \oe uvre de ces mécanismes au sein de la plate-forme
\noun{Jamus}. Le paragraphe~\ref{sec:related-work} présente quelques
travaux dont la démarche ou la finalité s'apparentent à la nôtre.  En
guise de conclusion, le paragraphe~\ref{sec:conclusion} résume les
différents éléments clés de notre approche et énumère quelques unes
des perspectives offertes par ce travail.


\section{\label{sec:overview}Vue d'ensemble de l'approche proposée}

Certains composants logiciels présentent des besoins en ressources
invariables et parfaitement identifiés, alors que d'autres exhibent
des besoins en ressources partiels qui évoluent en cours d'exécution
(e.g. composants dont les besoins en ressources sont liés aux choix de
l'utilisateur). Il nous paraît donc essentiel que les composants
logiciels puissent exprimer leurs besoins en ressources dynamiquement.
De la même manière, pour pouvoir tenir compte des besoins en
ressources exprimés par les composants logiciels, l'environnement de
déploiement des composants doit être doté de mécanismes lui permettant
de déterminer dynamiquement s'il est en mesure de satisfaire ces
besoins au vu des ressources dont il dispose, et doit être pourvu de
mécanismes lui permettant de gérer ces ressources afin d'en assurer la
disponibilité le cas échéant.

\begin{figure}
\begin{center}\includegraphics[%
  width=0.95\columnwidth]{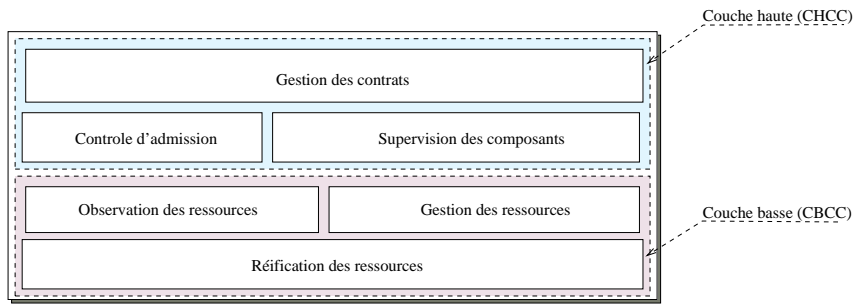}\end{center}

\caption{\label{fig:framework-overview}Fonctionnalités offertes par le cadre
de conception}
\end{figure}

Dans cette optique, nous avons conçu un cadre de conception qui
définit la structure d'un système de contractualisation dynamique des
ressources.  Ce cadre de conception, dont une vue d'ensemble est
présentée dans la figure~\ref{fig:framework-overview}, est organisé en
deux couches.  Chaque couche peut être complétée et étendue à volonté
par les développeurs, leur permettant ainsi de concevoir un système de
contractualisation répondant précisément à leurs besoins.

La couche inférieure du cadre de conception définit la structure d'un
ensemble de fonctionnalités conçues pour offrir une représentation
d'un environnement par réification des ressources de ce dernier, et
conçues pour observer et gérer ces ressources à travers cette
représentation.  Les fonctionnalités de la couche inférieure du cadre
de conception ont été définies indépendamment de celles de la couche
supérieure.  Elles peuvent donc par exemple être utilisées pour
concevoir des mécanismes de supervision et de contrôle de machines
hôtes ou des composants logiciels adaptatifs.

La couche supérieure du cadre de conception définit quant à elle la
structure d'une fonctionnalité de contrôle d'admission, qui permet à
un environnement de déploiement de déterminer s'il est en mesure de
satisfaire les besoins en ressources exprimés par les composants
logiciels; la structure d'une fonctionnalité de supervision des
composants, qui permet de vérifier que les composants utilisent les
ressources conformément à leur contrat; et la structure d'une
fonctionnalité qui offre le moyen de réifier les contrats sous la
forme objet, ainsi que de négocier et de vérifier dynamiquement ces
contrats.


\section{\label{sec:contrats}Contrats et  avenants}

\subsection{Contrats}

Les contrats que nous avons définis permettent d'exprimer les besoins
en ressources des composants logiciels de manière qualitative (e.g.
droits d'accès, contraintes de disponibilité) et quantitative (e.g.
quotas). Ces contrats sont définis comme un ensemble de profils
d'utilisation des ressources. Un profil est composé de quatre
attributs. Le premier attribut identifie une ressource de manière
précise. Le deuxième attribut et le troisième attribut définissent
respectivement les droits d'accès et les quotas requis par le
composant vis-à-vis de cette ressource. Le quatrième attribut spécifie
quant à lui la manière selon laquelle l'environnement de déploiement
doit assurer la disponibilité de cette ressource. Grâce à ce dernier
attribut, les composants logiciels peuvent préciser que certaines
ressources doivent leur être réservées, alors que d'autres ne doivent
pas nécessairement l'être. Ils peuvent ainsi indiquer à leur
plate-forme de déploiement comment ils sont susceptibles de se
comporter en cours d'exécution.

À titre d'exemple, considérons un composant logiciel baptisé
\emph{JMailer} qui offre le moyen de rédiger et d'envoyer des
courriers électroniques.  Ce composant offre à l'utilisateur deux
modes de fonctionnements possibles. Dans le premier mode,
l'utilisateur doit spécifier sans aucune assistance l'adresse des
destinataires, le sujet et le contenu du courrier électronique. Dans
le second mode, une assistance lui est fournie pour saisir les
adresses électroniques à travers l'utilisation d'un carnet d'adresses.
Pour pouvoir fonctionner dans le mode sans assistance, ce composant
logiciel exige de pouvoir lire et écrire jusqu'à 500 Kilo-octets de
données dans le répertoire \emph{\textasciitilde{}/.jmailer}, et
demande à disposer de 1 Méga-octet de mémoire. Pour s'exécuter dans le
second mode, le composant \emph{JMailer} demande à pouvoir lire et
écrire jusqu'à 500 Kilo-octets de données dans le répertoire
\emph{\textasciitilde{}/.jmailer} et jusqu'à 1 Méga-octets de données
dans le répertoire \emph{\textasciitilde{}/.jaddrbook}, et exige de
disposer de 2 Méga-octet de mémoire. Pour contractualiser ces
conditions d'accès aux ressources, notre composant de démonstration
\emph{JMailer} pourra par exemple soumettre à son environnement de
déploiement les contrats \emph{contract1} et \emph{contract2} de la
figure~\ref{fig:exemple-contrats}.

\begin{figure}[t]
\begin{flushleft}\hrule{}~\\
\textsf{\footnotesize ~}\textsf{\scriptsize ~~~~int Ko=1024;
int Mo=1024{*}1024;}\\
\textsf{\footnotesize ~}\textsf{\scriptsize ~~~~ResourceUtilisationProfile
r1, r2, r3, r4;}\\
\textsf{\footnotesize ~}\textsf{\scriptsize ~~~~}\textsf{\emph{\scriptsize //}}
\textsf{\scriptsize }\textsf{\emph{\scriptsize Accès en lecture et
écriture au répertoire \textasciitilde{}/.jmailer}}\\
\textsf{\footnotesize ~}\textsf{\scriptsize ~~~~}\textsf{\emph{\scriptsize //
quotas = 500 Koctets en lecture et écriture}}\\
\textsf{\footnotesize ~}\textsf{\scriptsize ~~~~}\textsf{\textbf{\scriptsize r1
= new}} \textsf{\scriptsize ResourceUtilisationProfile}\textsf{\textbf{\scriptsize (new}}
\textsf{\scriptsize FilePattern(\char`\"{}\textasciitilde{}/.jmailer\char`\"{}),}\textsf{\textbf{\scriptsize ~new}}
\textsf{\scriptsize FilePermission(FilePermission.ALL), }\\
\textsf{\textbf{\scriptsize ~~~~~~~~~~~~~~~~~~~~~~~~~~~~~~~~~~~~~~~~~~~~~~~~~~new}}
\textsf{\scriptsize FileQuota(500{*} Ko, 500{*}Ko),}\textsf{\textbf{\scriptsize ~new}}
\textsf{\scriptsize BestEffort())); }\\
\textsf{\scriptsize ~~~~~}\textsf{\emph{\scriptsize //}} \textsf{\scriptsize }\textsf{\emph{\scriptsize Accès
en lecture et écriture au répertoire \textasciitilde{}/.jaddressbook}}\\
\textsf{\emph{\scriptsize ~~~~~// quotas = 1 Moctets en lecture
et écriture}}\\
\textsf{\scriptsize ~~~~~}\textsf{\textbf{\scriptsize r2 = new}}
\textsf{\scriptsize ResourceUtilisationProfile}\textsf{\textbf{\scriptsize (new}}
\textsf{\scriptsize FilePattern(\char`\"{}\textasciitilde{}/.jaddrbook\char`\"{}),}\textsf{\textbf{\scriptsize ~new}}
\textsf{\scriptsize FilePermission(FilePermission.ALL), }\\
\textsf{\textbf{\scriptsize ~~~~~~~~~~~~~~~~~~~~~~~~~~~~~~~~~~~~~~~~~~~~~~~~~~new}}
\textsf{\scriptsize FileQuota(1{*} Mo, 1{*}Mo),}\textsf{\textbf{\scriptsize ~new}}
\textsf{\scriptsize BestEffort())); }\\
\textsf{\scriptsize ~~~~~}\textsf{\emph{\scriptsize //}} \textsf{\scriptsize }\textsf{\emph{\scriptsize réservation
de 1 Moctets de mémoire}}\\
 \textsf{\scriptsize ~~~~~}\textsf{\textbf{\scriptsize r3 = new}}
\textsf{\scriptsize ResourceUtilisationProfile}\textsf{\textbf{\scriptsize (new}}
\textsf{\scriptsize MemoryPattern(),}\textsf{\textbf{\scriptsize ~new}}
\textsf{\scriptsize MemoryPermission(), }\\
\textsf{\textbf{\scriptsize ~~~~~~~~~~~~~~~~~~~~~~~~~~~~~~~~~~~~~~~~~~~~~~~~~~~new}}
\textsf{\scriptsize MemoryQuota(1{*} Mo),}\textsf{\textbf{\scriptsize ~new}}
\textsf{\scriptsize ResourceReservation())); }\\
\textsf{\scriptsize ~~~~~}\textsf{\emph{\scriptsize //}} \textsf{\scriptsize }\textsf{\emph{\scriptsize réservation
de 2 Moctets de mémoire}}\\
 \textsf{\scriptsize ~~~~~}\textsf{\textbf{\scriptsize r4 = new}}
\textsf{\scriptsize ResourceUtilisationProfile}\textsf{\textbf{\scriptsize (new}}
\textsf{\scriptsize MemoryPattern(),}\textsf{\textbf{\scriptsize ~new}}
\textsf{\scriptsize MemoryPermission(), }\\
\textsf{\textbf{\scriptsize ~~~~~~~~~~~~~~~~~~~~~~~~~~~~~~~~~~~~~~~~~~~~~~~~~~~new}}
\textsf{\scriptsize MemoryQuota(2{*} Mo),}\textsf{\textbf{\scriptsize ~new}}
\textsf{\scriptsize ResourceReservation())); }\end{flushleft}{\scriptsize \par}

\begin{flushleft}\textsf{\scriptsize ~~~~~ResourceOrientedContract
contract1=} \textsf{\textbf{\scriptsize new}} \textsf{\scriptsize ResourceOrientedContract
(\{r1, r2, r4\});}\\
\textsf{\scriptsize ~~~~~ ResourceOrientedContract contract2=}
\textsf{\textbf{\scriptsize new}} \textsf{\scriptsize ResourceOrientedContract
(\{r1, r3\});}\\
~\\
\hrule{}\end{flushleft}

\caption{\label{fig:exemple-contrats}Exemple de contrats pouvant être soumis
à la plate-forme \noun{Jamus} }
\end{figure}

\subsection{Avenants}

Ainsi que nous l'avons mentionné précédemment, les besoins en
ressources des composants logiciels sont susceptibles d'évoluer au
cours du temps. Il est donc nécessaire d'offrir aux composants la
possibilité de pouvoir renégocier dynamiquement les modalités du
contrat qu'ils ont passées avec leur environnement de déploiement. Le
mécanisme d'avenants que nous avons introduit dans notre cadre de
conception offre  aux composants cette possibilité. En effet, ces
avenants sont définis comme un ensemble de clauses qui décrivent les
opérations d'ajout, de suppression, de modification de clause d'un
contrat. À l'instar des contrats, les avenants sont réifiés sous la
forme objet afin de pouvoir être définis et négociés dynamiquement par
les composants et leur environnement.

La figure~\ref{fig:exemple-avenant} présente un exemple d'avenant qui
peut être soumis par le composant \emph{JMailer} à son environnement
de déploiement pour renégocier les modalités du contrat qui les lie.
Cet avenant a pour effet d'ajouter un profil d'utilisation des
ressources portant sur les conditions d'accès au répertoire
\emph{/tmp} dans le contrat \emph{contract2} présenté dans la
figure\emph{~\ref{fig:exemple-contrats}.}

\begin{figure}
\begin{flushleft}\hrule{}~\\
\textsf{\scriptsize ~~~~~int Mo=1024{*}1024;}\\
\textsf{\scriptsize ~~~~~ResourceUtilisationProfile r5;}\\
\textsf{\scriptsize ~~~~~}\textsf{\emph{\scriptsize // accès
en lecture et en écriture au répertoire /tmp}}\\
\textsf{\scriptsize ~~~~~}\textsf{\textbf{\scriptsize r5 = new}}
\textsf{\scriptsize ResourceUtilisationProfile}\textsf{\textbf{\scriptsize (new}}
\textsf{\scriptsize FilePattern(\char`\"{}/tmp\char`\"{},),}\textsf{\textbf{\scriptsize ~new}}
\textsf{\scriptsize FilePermission(FilePermission.ALL), }\\
\textsf{\textbf{\scriptsize ~~~~~~~~~~~~~~~~~~~~~~~~~~~~~~~~~~~~~~~~~~~~~~~~~~new}}
\textsf{\scriptsize FileQuota(2{*} Mo, 2{*}Mo),}\textsf{\textbf{\scriptsize ~new}}
\textsf{\scriptsize BestEffort());}\\
\textsf{\scriptsize ~~~~~AmendmentClause ac1 =~}\textsf{\textbf{\scriptsize new}}
\textsf{\scriptsize AmendmentClause(AmendmentClause.ADD, r5);}\\
\textsf{\scriptsize ~~~~~Amendment a1=} \textsf{\textbf{\scriptsize new}}
\textsf{\scriptsize Amendment (contract2, \{ac1\});}\\
{\scriptsize ~}\\
\hrule{}\end{flushleft}

\caption{\label{fig:exemple-avenant}Exemple d'un avenant pouvant être soumis
à la plate-forme \noun{Jamus}}
\end{figure}

Le paragraphe suivant présente les mécanismes qui permettent de gérer
ces contrats et ces avenants de manière dynamique au sein de la
plate-forme \noun{Jamus}.


\section{\label{sec:jamus}La plate-forme \noun{Jamus}}

\noun{Jamus} (\emph{Java Accommodation of Mobile Untrusted Software})
est une plate-forme expérimentale dédiée à l'hébergement de composants
mobiles Java capables d'exprimer leurs besoins vis-à-vis des ressources.
\noun{Jamus} a été conçue par spécialisation de la couche supérieure
de notre cadre de conception. Elle s'appuie sur les fonctionnalités
d'observation et de contrôle d'accès des ressources offertes par l'environnement
d'exécution \noun{Raje} (\emph{Resource-Aware Java Environment}),
environnement que nous avons conçu par spécialisation de la couche
inférieure du cadre de conception. Cet environnement \noun{Raje}
n'est pas présenté plus amplement dans cet article par manque
de place. Le lecteur pourra toutefois se référer à~\cite{TheseNico}
pour avoir une présentation détaillée de \noun{Raje.}

La plate-forme \noun{Jamus} offre aux composants logiciels le moyen
de contractualiser dynamiquement leurs modalités d'accès aux ressources.
 En outre, étant dédiée à l'hébergement de composants logiciels potentiellement
non dignes de confiance, la plate-forme \noun{Jamus} soumet chaque
composant hébergé à une supervision constante au cours de son exécution
et interdit toute utilisation des ressources non conforme aux contrats
qu'ils ont souscrits avec elle.

\subsection{Gestion des contrats}

Dans \noun{Jamus}, les propositions de contrats soumises par les
composants logiciels sont évaluées par un courtier de ressources.  Ce
courtier implémente une fonctionnalité de contrôle d'admission et un
mécanisme de réservation des ressources afin d'offrir aux composants
hébergés une certaine qualité de service vis-à-vis de la disponibilité
des ressources. La figure~\ref{fig:contract-management} montre comment
ce courtier est interrogé par le gestionnaire de contrats de la
plate-forme \noun{Jamus} pour évaluer les contrats. Le processus de
contractualisation des ressources mis en \oe uvre au sein de la
plate-forme \noun{Jamus} comporte deux phases : une phase de
soumission des contrats et une phase de souscription des contrats. Ces
deux phases ont été distinguées afin de permettre aux composants
logiciels de pouvoir soumettre à leur environnement de déploiement
plusieurs contrats pour évaluation avant de souscrire un contrat
particulier avec celui-ci.

\subsubsection{Soumission des contrats}

Lors du démarrage de la plate-forme, le courtier reçoit en paramètre
un ensemble de profils d'utilisation des ressources décrivant les
ressources disponibles au sein de la plate-forme \noun{Jamus}. Au vu
de ces informations et de celles qu'il a maintenues à jour en fonction
des contrats souscrits avec les composants hébergés par la plate-forme
\noun{Jamus}, le courtier de ressources décide de la validité des
contrats soumis par les composants candidats à l'hébergement.  Un
contrat est déclaré acceptable par le courtier si toutes les clauses
de celui-ci peuvent être satisfaites.  Lorsqu'un contrat est déclaré
inacceptable par le courtier de ressources, le gestionnaire de
contrats de la plate-forme \noun{Jamus} invoque une nouvelle fois ce
courtier afin de connaître les profils qui ne peuvent pas être
satisfaits (invocation de la méthode \emph{getConflictingClause()}).
Sur la base des informations retournées par le courtier de ressources,
le gestionnaire de contrats construit un rapport de soumission qu'il
retourne au composant logiciel. Ces informations visent à aider ce
dernier dans la formulation de nouvelles propositions de contrats.

\begin{figure}
\begin{center}\includegraphics[%
  width=0.70\columnwidth,
  height=8cm]{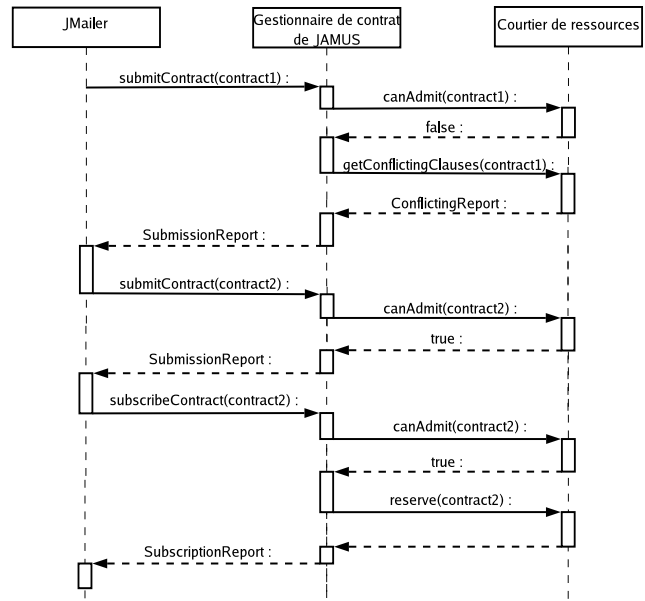}\end{center}

\caption{\label{fig:contract-management}Exemple de négociation des modalités
d'accès aux ressources entre le composant \emph{JMailer} et la plate-forme
\noun{Jamus}}
\end{figure}

La figure~\ref{fig:contract-management} illustre ces propos en
présentant un scénario de négociation de contrats entre notre
composant de démonstration \emph{JMailer} et la plate-forme
\noun{Jamus}.

\subsubsection{Souscription des contrats }

Lors de la phase de souscription, les contrats sont évalués une
nouvelle fois avant que les ressources ne soient réservées pour le
composant par le courtier. Cette réévaluation est nécessaire dans
\noun{Jamus} car toute ou partie des ressources requises par le
composant candidat à l'hébergement peuvent avoir été allouées aux
autres composants présents dans la plate-forme \noun{Jamus} au moment
de la souscription du contrat. Lorsque cette réévaluation est passée
avec succès, le courtier réserve les ressources pour le composant
logiciel.  Cette réservation s'effectue en prélevant sur les quotas
des profils d'utilisation des ressources concernés les valeurs de
quotas définis dans les profils du contrat.

À titre d'exemple, le contrat \emph{contract2} ayant été déclaré
acceptable par le courtier de \noun{Jamus}, le composant
\emph{JMailer} peut initier une phase de souscription du contrat
\emph{contract2} (voir figure~\ref{fig:contract-management}).  Ce
contrat est évalué une nouvelle fois par le courtier de ressources
afin de déterminer si les besoins requis par le composant peuvent
toujours être satisfaits. Dans le cas présent, nous supposerons qu'ils
le sont. Le gestionnaire de contrats de \noun{Jamus} demande alors au
courtier de ressources de réserver les ressources pour le composant
\emph{JMailer}.

\subsection{Supervision des composants logiciels}

Chaque composant hébergé par la plate-forme \noun{Jamus} s'exécute au
sein d'un conteneur. Le conteneur permet d'isoler un composant en lui
offrant son propre espace de nommage.  Ainsi, les objets créés par les
différents composants s'exécutant au sein de la plate-forme
\noun{Jamus} ne peuvent être partagés et un composant ne peut accéder
aux objets manipulés par d'autres composants.

Chaque conteneur intègre un registre de ressources qui est chargé de
référencer les ressources créées par le composant logiciel considéré,
ainsi qu'un moniteur d'appli\-ca\-tion capable d'extraire des contrats
les profils d'utilisation des ressources, et capable d'instancier en
conséquence des moniteurs de res\-sour\-ces qui seront chargés de
s'assurer que l'utilisation qui est faite des ressources demeure
conforme aux modalités d'utilisation définies par le contrat souscrit
par le composant. La plate-forme \noun{Jamus} intègre un type de
moniteurs de ressources spécifique pour chaque type de ressources
considéré dans \noun{Raje}. Chaque conteneur intègre également un
auditeur de registre de res\-sour\-ces (\emph{ResourceTracker}) qui
est chargé d'indiquer aux moniteurs de ressources quelles sont les
ressources qu'ils doivent superviser.

\begin{figure}
\begin{center}\includegraphics[%
  width=0.80\columnwidth]{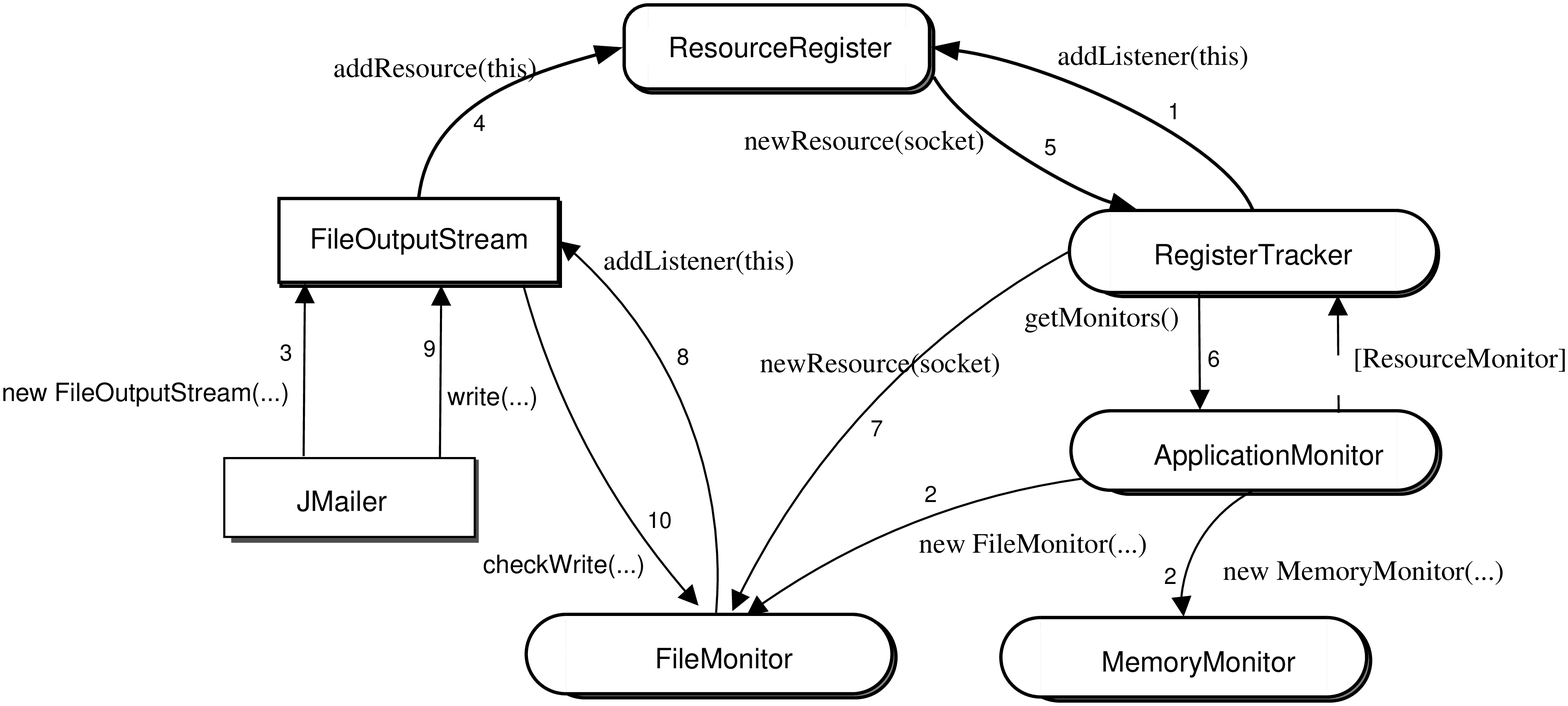}\end{center}

\caption{\label{fig:fonctionnement-conteneur}Principe de fonctionnement du
conteneur}
\end{figure}

Afin de présenter le principe de fonctionnement d'un conteneur,
considérons le scénario présenté dans la
figure~\ref{fig:fonctionnement-conteneur}.  Lors de la configuration
du conteneur, l'auditeur de registre de ressources s'enregistre en
tant qu'auditeur auprès du registre de ressources afin d'être informé
de toute nouvelle création ou destruction de ressources dans le
conteneur (action 1). Le moniteur d'application instancie quant à lui
des moniteurs de ressources en fonction des profils d'utilisation des
ressources contenus dans le contrat souscrit par le composant
considéré. Dans notre exemple, le moniteur d'application va instancier
un moniteur de type \emph{FileMonitor} afin de superviser les accès du
composant \emph{JMailer} au répertoire de l'utilisateur, ainsi qu'un
moniteur de type \emph{MemoryMonitor} pour vérifier l'utilisation qui
est faite de la mémoire (action 2). À l'issue de cette phase de
configuration, le conteneur charge le composant \emph{JMailer} et
initie son exécution.

Les objets ressources définis dans \noun{Raje} sont conçus de manière
à s'enregistrer spontanément auprès du registre de ressources lors
de leur création. Ainsi, l'objet ressource de type \emph{FileOutputStream}
est capable d'informer automatiquement le registre de ressources (action
3) lors de sa création (action 4). Le registre de ressources informe
à son tour l'auditeur de registre de ressources de la création de
cette nouvelle ressource (action 5). Ce dernier invoque alors le moniteur
d'application pour obtenir la liste des moniteurs de ressources instanciés
(action 6). Sur la base de cette information, l'auditeur du registre
de ressources sélectionne le moniteur devant superviser cette nouvelle
ressource (le moniteur \emph{FileMonitor}), et l'informe de l'apparition
de celle-ci dans le conteneur (action 7). Le moniteur \emph{FileMonitor}
s'enregistre alors en tant qu'auditeur auprès de la ressource \emph{FileOutputStream}
afin que cette dernière l'informe de toute action du composant \emph{JMailer}
(action 8). Ainsi, lorsque le composant écrit des données sur le flux
de données (action 9), celui-ci informe spontanément le moniteur \emph{FileMonitor}
de cette écriture (action 10).  Grâce à l'ensemble de ces mécanismes,
la plate-forme \noun{Jamus} est capable de superviser dynamiquement
les ressources utilisées par les composants logiciels qu'elle héberge.

Au vu de la complexité des mécanismes de supervision, on peut
légitimement penser que le surcoût de la supervision dynamique est
important. En fait, les évaluations des performances de la plate-forme
\noun{Jamus} que nous avons réalisées montrent que ce n'est pas le
cas, puisque dans le contexte d'observation le plus défavorable, le
temps d'accès à une ressource se trouve réduit d'environ 2\% par
moniteur.  L'ensemble de ces évaluations peuvent être consultées
dans~\cite{TheseNico}.

\subsection{Gestion de la violation des contrats}

La plate-forme \noun{Jamus} offre via les moniteurs d'applications et
les moniteurs de ressources le moyen de superviser les composants
logiciels au cours de leur exécution, ainsi que le moyen de
sanctionner les composants ayant violés les modalités de leur contrat.
Ces moniteurs s'appuient sur les fonctionnalités de contrôle d'accès
offertes par l'environnement \noun{Raje} pour sanctionner les
composants.

Nous avons identifié deux catégories de sanctions : les sanctions à
effet immédiat, et les sanctions à effet différé. Ces dernières sont
assujetties à un avertissement préalable qui vise à notifier le
composant responsable de la violation de contrat, et à lui permettre
ainsi de prendre les mesures nécessaires pour éviter l'application de
la sanction. Ces mesures peuvent impliquer par exemple la
renégociation des modalités du contrat ou la terminaison du contrat.
La figure~\ref{chap:jamus:sec:jamus:fig:exemple-sanction} présente ces
deux types de sanctions.  La première sanction est une sanction
différée s'appliquant au répertoire de l'utilisateur. Elle consiste à
rejeter les accès en lecture ou en écriture du composant à ce
répertoire au bout de deux violations successives des modalités
d'accès relatives à ce répertoire.  La seconde sanction est une
sanction de type immédiate qui porte sur les connexions TCP établies
sur le port 80. Cette sanction consiste à verrouiller l'accès du
composant à la socket concernée par la violation.

\begin{figure}
\begin{flushleft}\hrule{}~\\
 \textsf{\footnotesize }\textsf{\scriptsize ~~~~~Sanction s1,
s2;}\\
\textsf{\scriptsize ~~~~~}\textsf{\emph{\scriptsize //}} \textsf{\scriptsize }\textsf{\emph{\scriptsize Sanction
différée s'appliquant au répertoire \textasciitilde{}/}}\\
\textsf{\emph{\scriptsize ~~~~~// l'accès à la ressource sera
rejeté au bout de 2 violations successives}}\\
 \textsf{\scriptsize ~~~~~}\textsf{\textbf{\scriptsize s1 = new}}
\textsf{\scriptsize DifferedSanction(}\textsf{\textbf{\scriptsize new}}
\textsf{\scriptsize FilePattern(\char`\"{}\textasciitilde{}/\char`\"{}),
Sanction.REJECT, 2); }\\
 \textsf{\emph{\scriptsize ~~~~~// Sanction immédiate s'appliquant
aux connexion TCP sur le port 80}}\\
\textsf{\scriptsize ~~~~~}\textsf{\textbf{\scriptsize s2 = new}}
\textsf{\scriptsize ImmediateSanction}\textsf{\textbf{\scriptsize (new}}
\textsf{\scriptsize SocketPattern(\char`\"{}{*}\char`\"{}, 80), Sanction.LOCK);}{\scriptsize }\\
{\scriptsize ~}\\
\hrule{}\end{flushleft}

\caption{\label{chap:jamus:sec:jamus:fig:exemple-sanction}Exemple de sanctions
pouvant être utilisées pour configurer la plate-forme \noun{Jamus}}
\end{figure}


\section{\label{sec:related-work}Travaux apparentés}

\subsection{Travaux apparentés relatifs à la qualité de service}

La prise en compte des besoins en ressources des composants logiciels
et plus généralement des programmes d'application est une
problématique connue dans le domaine de la qualité de service.
Certains travaux se focalisent uniquement sur l'expression de la
qualité de service (e.g. QML~\cite{FrolundKoistinen00}), alors que
d'autres proposent à la fois des solutions pour exprimer la qualité de
service et transposer celle-ci en une expression des besoins en
ressources (e.g.
QuO~\cite{SchantzLoyallRodriguesSchmidt:middleware03},
QosTalk~\cite{WichadakulNahrstedt:cd02},
2k~\cite{WichadakulNahrstedtGuXu:middleware01}).  Cependant, ces
travaux proposent uniquement une expression quantitative et globale
des besoins en ressources des programmes d'application.  Bien qu'étant
parfaitement suffisante pour assurer la disponibilité des ressources
pour les applications, cette expression des besoins en ressources
n'est pas assez précise pour pouvoir être également utilisée au sein
des mécanismes de sécurité. Pourtant à l'instar de la plate-forme
\noun{Jamus}, il est possible de confronter les besoins exprimés par les
composants à la politique de sécurité mise en \oe uvre au sein de la
plate-forme de déploiement afin de déterminer si le composant peut
être déployé, et utiliser les besoins ainsi exprimés pour superviser
les composants en cours d'exécution.

\subsection{Travaux apparentés relatifs à la sécurité}

L'environnement d'exécution de Java~(JRE \textit{: Java Runtime
  Environment}) met en \oe uvre un modèle de sécurité connu sous le
nom de \emph{SandBox}.  Depuis la plate-forme Java~2, ce modèle de
sécurité repose sur la notion de domaine de
protection~\cite{JavaSecurityFutureAndPresent}.  Un domaine de
protection constitue un environnement d'exécution dont la politique de
sécurité peut être spécifiée sous la forme de permissions.  Le modèle
de sécurité mis en \oe uvre dans le JRE repose sur des mécanismes <<
~sans état~ >>. Il n'est donc pas possible d'imposer des contraintes
quantitatives sur les ressources manipulées au sein d'un domaine de
protection. En conséquence, les dysfonctionnements résultant de
l'utilisation abusive des ressources ne peuvent pas être
prévenus.

Des environnements tels que JRes~\cite{JRES} et
KaffeOS~\cite{ProcessesInKaffeOS} apportent un début de réponse au
problème du contrôle quantitatif des ressources posé par les
environnements Java traditionnels. Ils offrent en effet des mécanismes
permettant de comptabiliser --~et, éventuellement, de limiter~--
l'utilisation de chaque type de ressources par une entité active~(un
\emph{thread} dans le cas de JRes et un processus dans le cas de
KaffeOS).  Cependant dans ces environnements, les ressources sur
lesquelles portent la comptabilisation et le contrôle sont des
ressources globales. On peut ainsi comptabiliser les accès au réseau
réalisés par un \emph{thread}~(ou processus), mais on ne peut pas
distinguer les transmissions réalisées vers une machine donnée, ou
vers un numéro de port précis.

Les projets Naccio~\cite{NaccioPDH} et
Ariel~\cite{ArielProvidingFine-GrainedAccessControlForJavaPrograms}
proposent chacun un langage et des mécanismes permettant de définir de
manière très précise la politique de sécurité qui doit être appliquée
à un programme d'application lors de son exécution.  L'application
d'une politique de sécurité est réalisée statiquement, par réécriture
du \emph{bytecode} du programme d'application, mais aussi des classes
de l'API de la plate-forme Java.  Cette approche se prête donc bien à
la génération anticipée d'un ensemble d'API pré-définies garantissant
chacune le respect d'une politique de sécurité générique.  En
revanche, elle ne permettrait pas, comme le fait la plate-forme
\noun{Jamus}, d'assurer la supervision d'un programme d'application en
fonction d'une politique de sécurité définie à partir des besoins
exprimés par ce même programme lors de son démarrage.


\section{\label{sec:conclusion}Conclusion}

Dans cet article nous avons présenté un cadre de conception pour la
construction d'un système de contractualisation dynamique des
ressources, ainsi qu'une plate-forme d'accueil dédiée à l'hébergement
de composants logiciels qui a été mise en \oe uvre à l'aide de ce
cadre de conception. Cette plate-forme nous a permis de mettre en
évidence la pertinence et la validité de notre approche. Nous étudions
actuellement la possibilité d'intégrer notre approche au sein de
modèles de composants logiciels existants (e.g. Fractal, OSGi) afin
d'en améliorer les mécanismes de déploiement.

\vspace{-0.5cm}
\bibliography{decor-2004}

\end{document}